\documentclass[twocolumn,secnumarabic,amssymb, nobibnotes, aps, prd]{revtex4-2}
\newcommand{\figref}[1]{Fig.\ref{#1}}
\newcommand{\figsref}[1]{Figs.\ref{#1}}
\usepackage{amsmath}
\usepackage{newtxtext,newtxmath}
\usepackage{braket}
\usepackage[dvipdfmx,hiresbb]{graphicx}
\usepackage{color}

\begin{document}

\title{Coherent beam splitting of flying electrons driven by a surface acoustic wave}%

\author{R.Ito$^1$}
\author{S.Takada$^2$}
\author{A.Ludwig$^3$}
\author{A.D.Wieck$^3$}
\author{S.Tarucha$^1$}
\author{M.Yamamoto$^1$}

\affiliation{$^1$Center for Emergent Matter Science, RIKEN, 2-1 Hirosawa, Wako, Saitama 351-0198, Japan}
\affiliation{$^2$National Institute of Advanced Industrial Science and Technology, National Metrology Institute of Japan, 1-1-1 Umezono, Tsukuba,Ibaraki 305-8563,Japan}
\affiliation{$^3$Angewandte Festk\"{o}rperphysk, Ruhr-Universit\"{a}t Bochum, D-44780 Bochum Germany}

\begin{abstract}
We develop a coherent beam splitter for single electrons driven through two tunnel-coupled quantum wires by surface acoustic waves (SAWs). The output current through each wire oscillates with gate voltages to tune the tunnel-coupling and potential difference between the wires. This oscillation is assigned to coherent electron tunneling motion that can be used to encode a flying qubit and is well reproduced by numerical calculations of time evolution of the SAW-driven single electrons. The oscillation visibility is currently limited to about 3\%, but robust against decoherence, indicating that the SAW-electron can serve as a novel platform for a solid-state flying qubit.

\end{abstract}
\maketitle

In quantum optics, quantum information is encoded on photons, called flying qubits. The photon qubits are usually applied for implementing quantum communication but not quantum computation because of the scalability problem. A new approach of sorting out this problem has recently been proposed in which a qubit array can be stored in a loop of an optical channel and universal operations can be achieved by connecting only a few fundamental physical gates to the optical channel\cite{PhysRevLett.119.120504, Takedaeaaw4530}. Such an architecture of the photon flying qubits is different from those of solid-sate qubits that require the physical gate structures to be scalable. On the other hand, similar to the optical systems, quantum circuits of electrons propagating through one-dimensional (1D) wires are also able to host electrons as flying qubits. In previous studies, single flying qubit manipulation has been demonstrated in electronic Mach-Zehnder interferometers\cite{Ji:2003fk,Yamamoto:2012fj}. However, the flying qubits in this study consist of electrons continuously injected from the static macroscopic reservoirs, and therefore the electron wave functions are spatially spread along the longitudinal direction. Thus, these qubits are incompatible with the photon qubit arrays in quantum optics.

On the other hand, various on-demand sources of finite-size single electrons that resemble the photon arrays have recently been demonstrated\cite{Feve:2007hb,Dubois:2013fk,Jullien:2014uq,Bisognin:2019aa, McNeil:2011aa,Hermelin:2011fk,BertrandB.:2016qy,Takada:2019aa}. One of the ways is to  use a surface acoustic wave (SAW) traveling along a depleted 1D channel made in a piezoelectric medium. The SAW generates a moving electrostatic potential (MQD: moving quantum dot) to capture a single electron from a reservoir or a static quantum dot and transport it, while confining it in the SAW potential minimum\cite{0953-8984-8-38-001,PhysRevB.56.15180}. The spatial extension of each trapped single electron wave function can be made smaller than 100 nm. In a train of such single electrons, electrons are 1 $\mu$m apart from each other. This spacing is much smaller than the optical qubit spacing, and allows us to create a qubit array, encoded on a train of single electrons. Therefore the SAW-driven electrons can be a promising candidate to construct a scalable quantum computing system, resembling the photon arrays.

Among the required functions for flying-electron-based quantum computing, on-demand single electron emission and detection for SAW-driven electrons have already been achieved with high efficiency of 99\%\cite{Takada:2019aa}. It has also been shown that the spin information of the SAW-driven single electron is preserved while being transferred between distant quantum dots through a depleted 1D channel\cite{BertrandB.:2016qy}. It is also possible to manipulate its spin using the spin-orbit interaction during transportation\cite{Sanada:2013fk}. One of the remaining challenges to implement the flying electron qubits is now the quantum manipulation of motion, i.e. orbital state, of a SAW-driven single electron. Especially a fully tunable coherent beam splitter is a key ingredient in the flying qubit operation; however, its demonstration has still been elusive owing to technical difficulties.

The coherent beam splitter for SAW-driven single electrons was proposed more than a decade ago\cite{KATAOKA2006546}. It consists of two tunnel-coupled parallel quantum wires\cite{doi:10.1063/1.102657,doi:10.1063/1.102877,PhysRevLett.84.5912} which is similar to those used in previous studies on static electron Mach-Zehnder interferometers\cite{Ji:2003fk,Yamamoto:2012fj}, and the qubit state is encoded by electron occupation of either of the tunnel-coupled wires (TCWs). Although there are some reports demonstrating splitting or directional control of the electron flow in a similar device\cite{:/content/aip/journal/apl/88/8/10.1063/1.2176847,KATAOKA2006546,KATAOKA20081017,Takada:2019aa}, coherent tunneling of SAW-driven electrons propagating through the TCWs has never been addressed so far.

\begin{center}
\begin{figure*}
	\includegraphics[width=17.8cm]{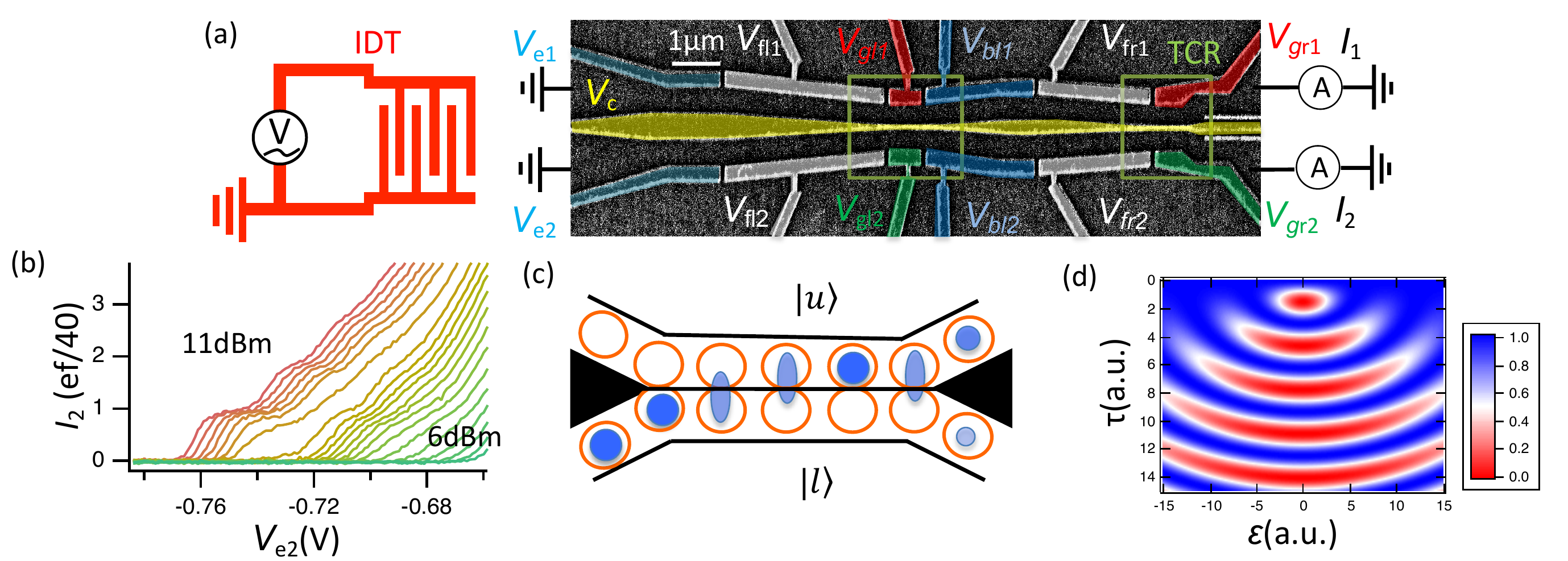}
	\caption{(a) Scanning electron microscope image of TCWs. Two TCRs between TCWs are indicated by the green square flames. IDT is placed 1.3 mm away from the TCWs to the left. SAW generated by the IDT carries electrons to the TCR. (b) Quantized current observed when SAW-driven single electrons are injected to the left bottom wire defined by gate voltages of $V_{\rm c}$ and $V_{\rm e2}$. $V_{\rm e2}$ is only changed with keeping $V_{\rm c}$ constant at -1 V. The microwave power applied on the IDT is varied from 11 dBm (red) to 6 dBm (green) in steps of 0.25 dBm.(c) Schematic of the TCR with propagation of SAW driven single electrons. The qubit is encoded by electron propagation in either side of the wire. (d) Calculation of electron tunnel oscillation pattern (output current of the lower channel normalized by the total current) as a function of the tunneling energy ($\tau$) and detuning ($\epsilon$) using a two-site Hubbard model. \label{dev2}}
\end{figure*}
\end{center}

Here, we realize the coherent beam splitting of SAW-driven single electrons in the TCWs. The SAW potential loads single electrons from a reservoir to its potential minima to construct an array of single electrons that are transported with a fixed time interval. We measure two output currents obtained for the SAW-driven electrons passing the TCWs and find them consistent with numerical calculations performed by solving the time-dependent Schr\"{o}dinger equation. The visibility of the coherent beam splitting we obtained is low, but robust against increase in temperature. The low visibility can be attributed to randomness of the initial electron state at the entrance of the TCW. To the best of our knowledge, this is the first demonstration of a coherent electron beam splitter, which serves as a milestone towards the realization of on-demand single electron quantum optical devices.

The device used in our experiment is made out of GaAs/AlGaAs heterostructure which contains a two-dimensional electron gas (2DEG) with a mobility of $1.5\times 10^6 {\rm cm^2/Vs}$ and carrier density of $1.46\times 10^{11} {\rm cm}^{-2}$ cooled to low temperatures without bandgap-photon illumination. Two TCWs are defined by using a Schottky gate technique and they are only tunnel coupled in two separate tunnel coupled regions (TCRs) (see two square frames in \figref{dev2}a). Both wires are depleted to isolate the SAW-driven electrons from the electrons in the surrounding. An interdigital transducer (IDT) that converts microwaves to SAWs is placed 1.3 mm away from the entrance of the 1D wires. The period of the IDT fingers, i.e., the SAW wavelength is $1\mu {\rm m}$, while the number of the IDT periods is 100. A 13-dBm microwave is applied to the IDT with 1/40 duty cycle to avoid heating of the entire device\cite{Utko:2006aa,:/content/aip/journal/apl/89/12/10.1063/1.2346372}. All gate electrodes and the IDT are fabricated by depositing Ti/Au with thicknesses of 15 nm/25 nm on the substrate surface. 

The SAW-driven electronic current through the wire is quantized in units of $ef$ times the duty cycle of 1/40, where $e$ is the elementary charge and $f$  the SAW frequency (\figref{dev2}b). We use the quantized SAW current as an electron source to the TCWs. The number of electrons transported in each MQD is then one or less. The SAW-driven electrons are loaded into the MQD and transported through the lower wire to one of TCRs at the SAW velocity of 2.7 km/s . Gate voltages applied to deplete the lower wire are adjusted such that the potential slope along the wires is slightly upward in the transport direction until electrons reach the TCR, thus protecting the electron from dropping forward the MQD.

We use the two-site Hubbard model to describe the electron state (flying qubit) evolution in the TCR. The Hamiltonian is given by
\begin{equation}
H=\frac{\epsilon}{2}\ket{l}\bra{l}-\tau\ket{u}\bra{l}-\tau\ket{l}\bra{u}-\frac{\epsilon}{2}\ket{u}\bra{u},
\end{equation}
where $\tau$ is the inter-channel tunnel-coupling energy and $\epsilon$ represents the detuning, which can be defined as the onsite energy difference between the two wires. The flying qubit state is encoded based on whether the electron trapped by the MQD is in the upper($\ket{u}$) or lower($\ket{l}$) wire (\figref{dev2}c). The time evolution of the electron state is represented as $\ket{\phi(t)}= e^{\frac{-iHt}{\hbar}}\ket{\phi(0)}$ with $\ket{\phi(0)}=\ket{l}$. The electron periodically oscillates between the two MQDs by tunneling through the center barrier during the time evolution. The probability of an electron flowing out of the lower wire is calculated as a function of $\epsilon$ and $\tau$, and shown in \figref{dev2}d. The final qubit state is then determined by measuring the output current in the TCWs. 

In our experiment, coherent inter-wire tunneling of SAW-driven electrons is investigated by sweeping the side gate voltages $V_{\rm gr1}$ and $V_{\rm gr2}$ ($V_{\rm gl1}$ and $V_{\rm gl2}$) for the right (left) TCR (see \figref{dev2}a). These voltages can be used to simultaneously tune both $\tau$ and $\epsilon$ for a fixed center-gate voltage $V_{\rm c}$. The difference between the side gate voltages, $V_{\rm gdr}=V_{\rm gr1}-V_{\rm gr2}$ ($V_{\rm gdl}=V_{\rm gl1}-V_{\rm gl2}$) is used as a control parameter for $\epsilon$, while their average, $V_{\rm gsr} = \frac{V_{\rm gr1}+V_{\rm gr2}}{2}$ ($V_{\rm gsl} = \frac{V_{\rm gl1}+V_{\rm gl2}}{2}$), to modify the coupling energy $\tau$ for the right (left) TCR. Even though there are two TCRs in the device, only one of them is adjusted to have an appropriate tunnel coupling in the experiment. The other is tuned to have two wires isolated i.e. their tunnel coupling is suppressed to zero. 

\begin{figure}
	\includegraphics[width=8.6cm]{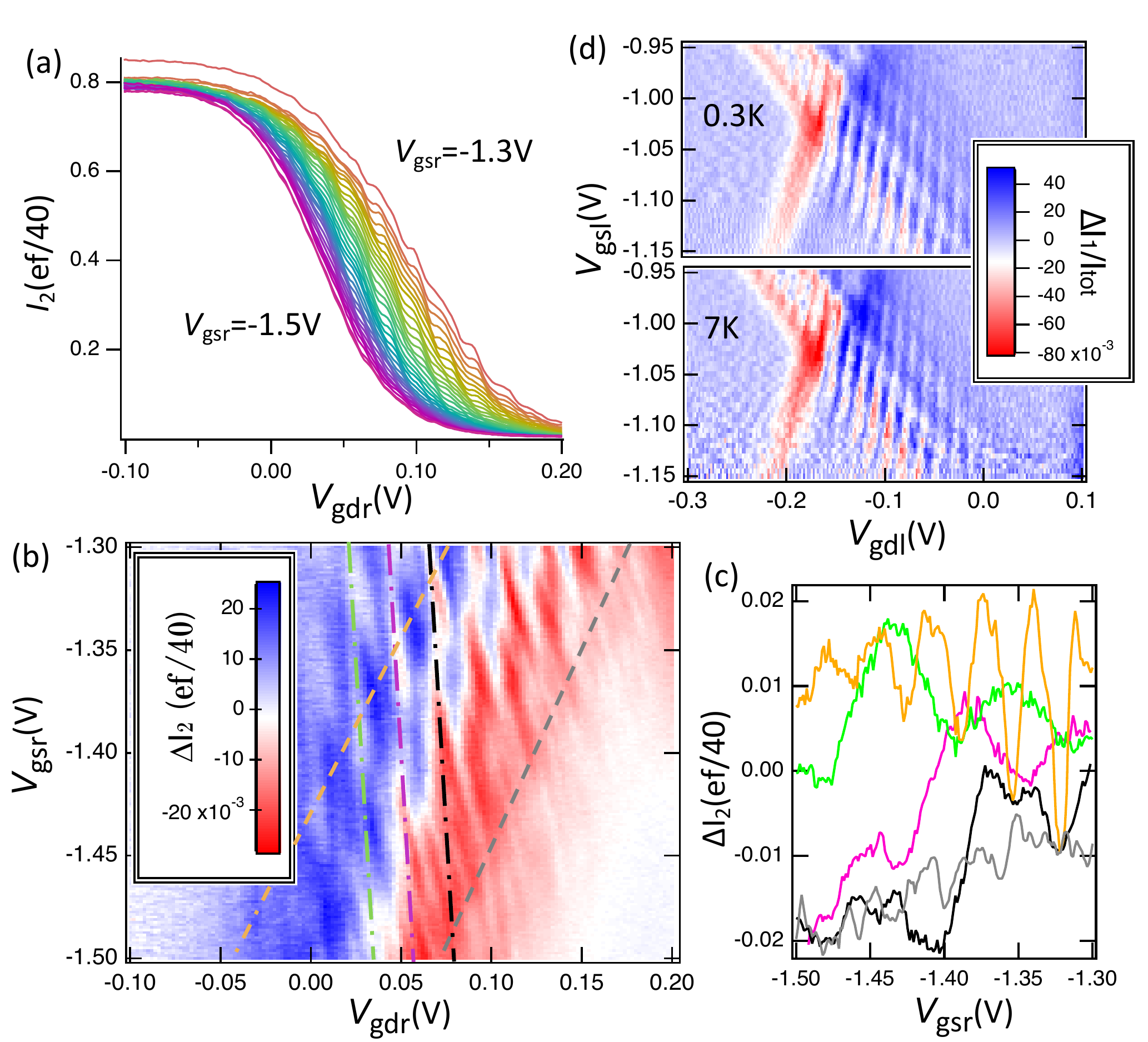}
	\caption{(a) Output current $I_2$ measured at the lower ohmic contact as a function of $V_{\rm gdr}=V_{\rm gr1}-V_{\rm gr2}$. The right TCR in \figref{dev2}a is used for the experiment. Electrons are injected from the lower wire. Line colors indicate different $V_{\rm gsr}=\frac{V_{\rm gr1}+V_{\rm gr2}}{2}$ values ranging from -1.3 V (red) to -1.5 V (purple) with $V_{\rm c}$=-0.7 V and $V_{\rm fr1} =V_{\rm fr2}$=-1.3 V. (b) The oscillating component of the current obtained by subtracting the smoothed background from the raw data in (a). (c) Intensity plot of (b) along the dashed lines and dot-dashed lines. $V_{\rm gdr}$ is simultaneously changed to trace the respective lines as $V_{\rm gsr}$. The line color setting corresponds to that in (b).
(d) Normalized oscillation component($\Delta I_1$) by the total current($I_{\rm tot}$) measured at 0.3 K and 7 K. For this experiment the left TCR is used and electrons are injected from the upper wire. $V_{\rm c}$=-0.7 V, $V_{\rm fl1} =V_{\rm fl2}$=-1.2 V and $V_{\rm bl1} =V_{\rm bl2}$= -1.3V.\label{tun2}}
\end{figure}

\figref{tun2}a shows the current $I_2$ measured at the output contact of the lower channel of the right TCR as a function of $V_{\rm gdr}$ and $V_{\rm gsr}$. The current $I_1$ at the other output varies simultaneously such that the total current $I_{\rm tot}=I_1+I_2$ is constant ($I_1$ and $I_{tot}$ are not shown). $I_{tot}$ is approximately $0.8ef/40$, less than the quantized value probably because there is a finite probability of an electron escaping from the MQD and being backscattered while traveling through the long entire depleted 1D wires. For $V_{\rm gdr}$ < -0.05 V, $I_2$=$I_{tot}$, indicating that all electrons flow through the lower wire. On the other hand as $V_{\rm gdr}$ is made more positive, $I_2$ becomes gradually smaller with accompanying ripples and finally quenched for $V_{\rm gdr}$> 0.2 V, indicating that all electrons flow through the upper wire. The ripple-like structure can only be observed when $I_2$ is close to half of the total current. Note that the ripple structure is absent when the number of electrons in each MQD is increased over one (see Supplementary information). To highlight the ripple structure, we subtract the background derived by smoothing the raw data along $V_{\rm gdr}$ and plot the outcome ($\Delta I_2$) as a function of $V_{\rm gdr}$ and $V_{\rm gsr}$ in \figref{tun2}b. Though the oscillation pattern does not resemble that derived from the simple two-level model (\figref{dev2}d), we find that a zoomed-in view of the pattern in \figref{tun2} shows a similarity as explained below.

In \figref{tun2}b we observe current oscillations in two directions as indicated by the dot-dashed and dashed lines, respectively. We here take the higher-lying orbital state in each MQD into account to explain these oscillations based on knowledge from the numerical simulation (explained later). Along the dot-dashed lines one of the higher-lying orbital states in the upper MQD and the initially loaded state in the lower MQD are energetically aligned. The tunnel-coupling energy $\tau$ between the MQD states changes with $V_{\rm gsr}$, causing the current to oscillate. Since $\tau$ only gradually changes with respect to $V_{\rm gsr}$ compared to $\epsilon$ with respect to $V_{\rm gdr}$, the WiFi-symbol-like pattern in \figref{dev2}d is squeezed horizontally to become like the oscillation along the dot-dahsed line. On the other hand along the dashed line, different higher-lying orbital states in the upper MQD are sequentially aligned with the initially loaded state in the lower MQD. $\tau$ between the MQD states is tuned to be constant along the dashed lines, and thus $I_2$ becomes small every time the MQD states align, providing current oscillation. \figref{tun2}c shows the intensity plot along the dot-dashed and dashed lines in \figref{tun2}b. The maximum visibility of the current oscillation obtained is about 3\% for the dot-dashed line in yellow and 2\% for the dashed line in green line, respectively.

To support the above described explanation and study inherent problems for realizing a coherent beam splitter we numerically simulate the electron motion in the TCR. The TCR potential profile is calculated for the gate geometry similar to the left TCR in \figref{dev2}a by solving Laplace's equation using a finite element method (see Supplementary Information). Here gate metal electrodes and electron reservoirs are only considered in the calculation. In addition, a constant dielectric constant of GaAs is assumed for the entire semiconductor. The electron motion is then numerically calculated based on the Cayley's form technique\cite{PhysRevE.62.2914} using a finite difference method for space and time. 
We assume that amplitude of the SAW-induced moving potential is 20 mV as obtained experimentally using the method described by Fletcher et al.\cite{PhysRevB.68.245310}; and that this value is not affected by the presence of the surface gates\cite{PhysRevB.58.10589,PhysRevB.74.035308,Takasu_2019}. The initial electron state is assumed to be in the ground state confined by the SAW and gate induced electrical potential.

\begin{figure}
	\includegraphics[width=8.6cm]{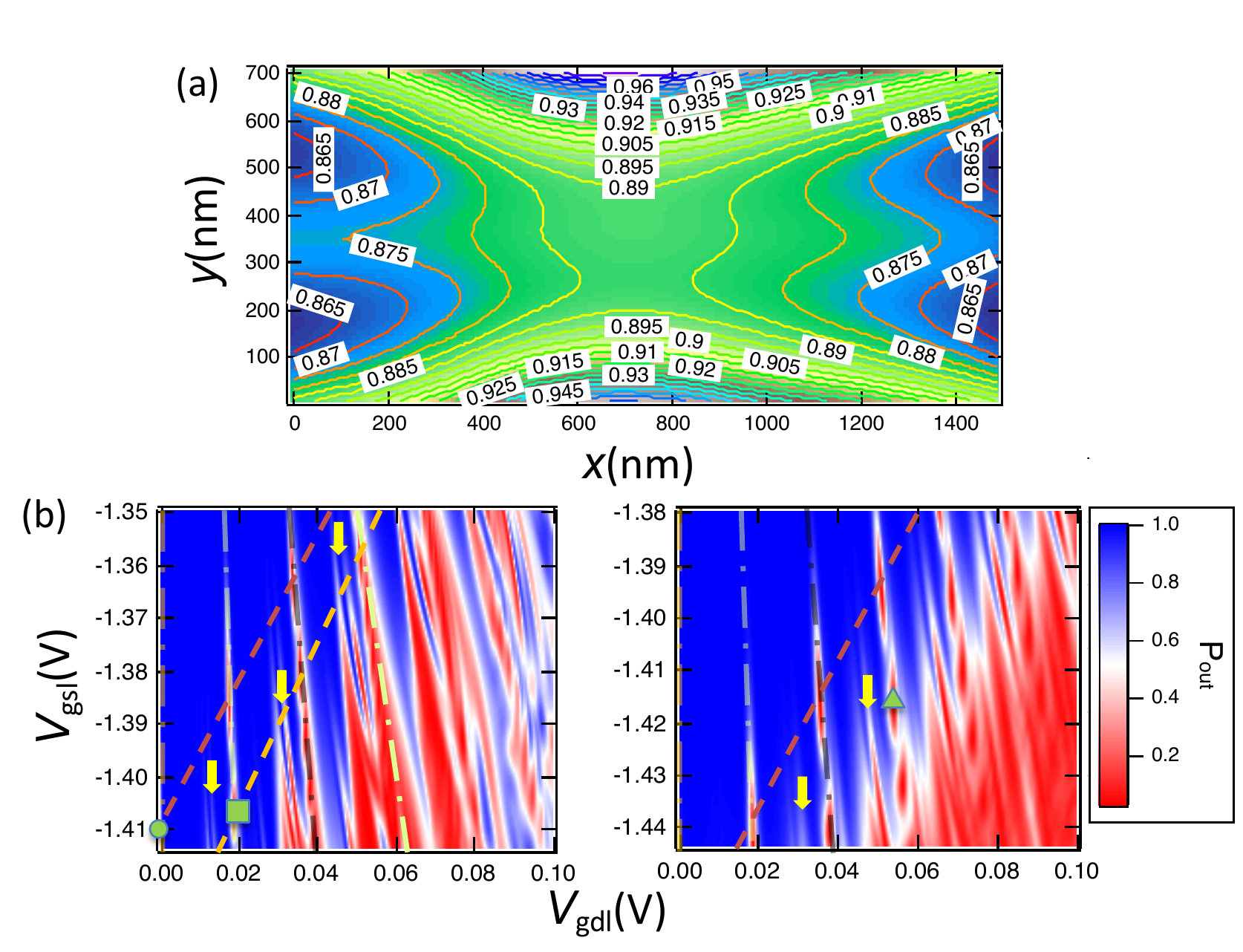}
	\caption{(a) Example of finite element method potential calculation for the gate structure of the left TCR in \figref{dev2}a. $V_{\rm c}=-1 V$,$V_{\rm fl1} =V_{\rm fl2}$=-1.2 V,$V_{\rm bl1} =V_{\rm bl2}$=-1.3 V,$V_{\rm gl1}$=-1.38 V, and $V_{\rm gl2}$=-1.42 V are assumed. (b) Numerically calculated probability $P_{\rm out}$ of electron staying in the lower wire through the TCR. Color scale shows $P_{\rm out}$ ranging from 0 to 1. Left: Typical result when electrons are fully trapped by the SAW potential during the time evolution. $V_{\rm c}$=-1 V,$V_{\rm fl1} =V_{\rm fl2}$=-1.1 V,$V_{\rm bl1} =V_{\rm bl2}=$=-1.2 V. Right: typical result when a party of electrons drop off the SAW potential at the middle of the TCR. $V_{\rm c}$=-1 V,$V_{\rm fl1} =V_{\rm fl2}$=-1.1 V, $V_{\rm bl1} =V_{\rm bl2}$=-1 V. \label{num}}
\end{figure}

The calculated results for the potential profile of the TCR and the probability $P_{\rm out}$ of the SAW-driven electron, staying in the same 1D wire, are shown in \figref{num}a and b, respectively. The calculated $P_{\rm out}$ reproduces well the experimentally observed features, i.e. two families of current oscillations along the dot-dashed and dashed lines in \figref{tun2}b. We figure out the origin of the two current oscillation families by analyzing the time evolution of SAW-driven electron at fixed gate voltages. 

\begin{figure}
	\includegraphics[width=8.6cm]{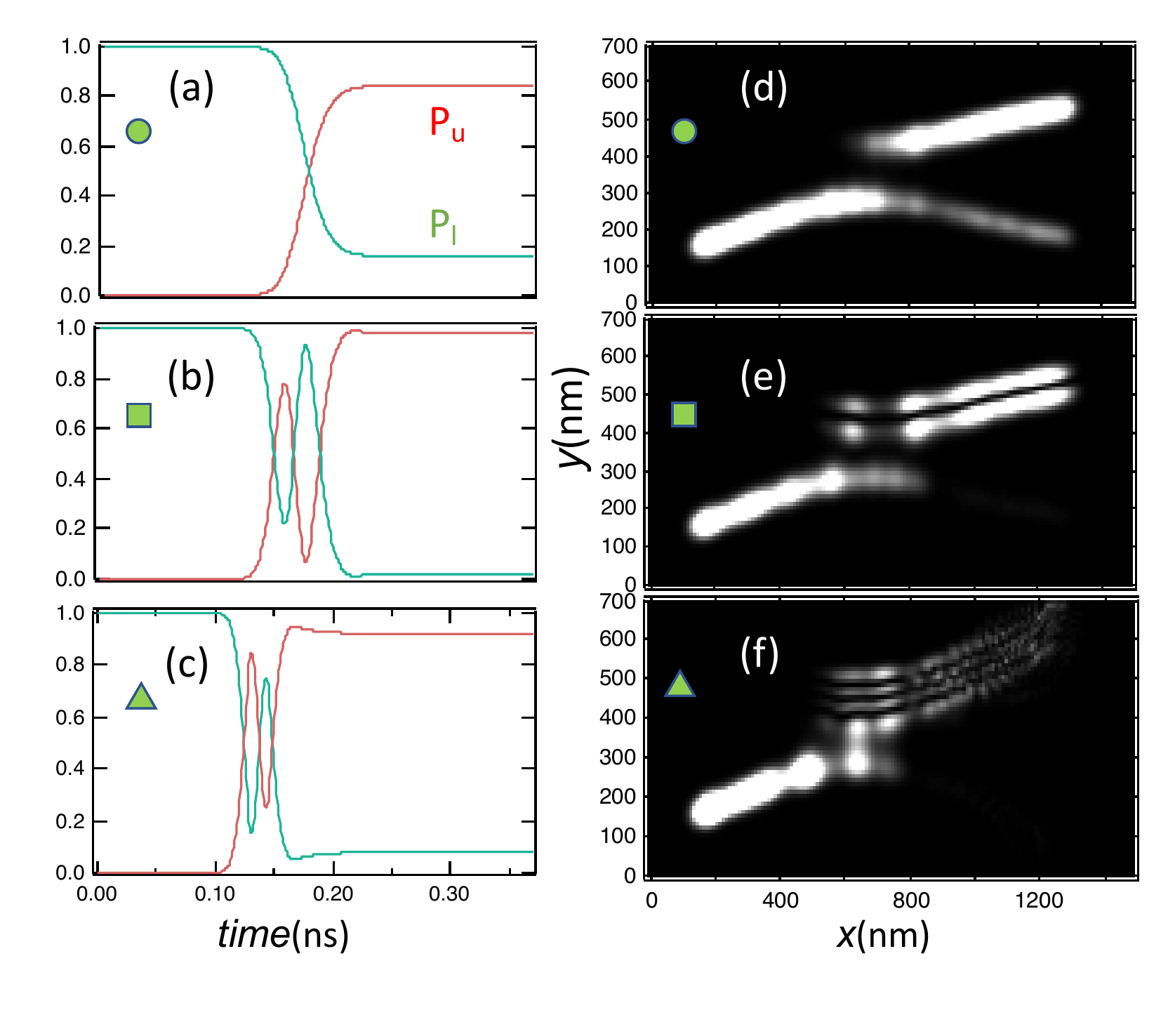}
	\caption{Calculated temporal and spatial distribution of an electron in the TCW for each marked point in \figsref{num}b and c. (a)--(c): Probability of finding an electron in the lower channel, $P_{\rm l}$ in green and upper channel, $P_{\rm u}$ in red, respectively at each time instance. The electron is injected from the TCW lower channel. (d)--(f): Cumulative SPD calculated for a potential profile as indicated in \figref{num}a. The distributions obtained at the time intervals of 15.4 ps are integrated to visualize the trajectory of a single electron. \label{move}}
\end{figure}

\figref{move} shows the calculated time evolution of the electron state at each gate voltage marked by the green circle, square and triangle in \figref{num}b, respectively. \figref{move}a--c show the probability $P_{\rm l} (P_{\rm u})$ of finding the electron in the lower (upper) wire by the green (red) line. On the other hand, \figref{move}d--f show the accumulation of 24 datasets of the spatial probability distribution (SPD) at fixed time intervals of 15.4 ps. Over the 24 datasets, the SAW travels by $1 \mu$m. These plots show the trajectory of electrons and shapes of the wave function in the transverse direction, and therefore profile the orbital states that contribute to the inter-wire tunneling of the SAW-driven single electrons. For example the green square is placed on the crossing of the second leftmost dot-dashed line and second topmost dashed line in the left panel of \figref{num}b. At this point, the ground state of the lower MQD having no node in SPD and the first excited state of the upper MQD having one node along the transverse direction are in resonance (see \figref{move}e). Note that it is not explicitly shown but the electron is in the ground state in each MQD along the longitudinal direction. Resonance of the ground and first excited state is maintained on the second leftmost dot-dashed line, changing the frequency of tunnel oscillation, i.e. number of electron tunneling. At the green square point, the SAW-driven electron undergoes three times tunneling between the wires as shown in \figref{move}b. The oscillation number is fixed on the second topmost dashed line. The same rule applies to the lines running through the green circle: The ground state is aligned between the upper and lower MQD and the electron tunnels only once from the lower to upper MQD as shown in \figref{move}a. 

Finally we address the case where $V_{\rm sb}$ is set to be more positive (right panel of \figref{num}b). In this situation, electrons drop off the MQD forward at the end of the TCR. The time evolution of $P_{\rm l}$, $P_{\rm u}$ and SPD at the triangle mark are plotted at \figref{move}c and f. In \figref{move}f, SPD is suppressed at x>800nm because the electron quickly escapes from the MQD. This results in a more distinct current oscillation along the dot-dashed line in \figref{num}b right, because a superposition state at the end of the TCR does not adiabatically fall into a local state ($\ket{u}$ or $\ket{l}$) in one of the two wires. We cannot determine whether an electron is trapped in or dropped from the MQD in experiment with right TCR such as \figref{tun2}b. However, with left TCR we observe that the tunneling signal becomes clearer by increasing $V_{\rm bl1}$ and $V_{\rm bl2}$, probably due to the reduced influence of above-mentioned adiabatic state evolution (see supplementary information). 

The visibility of current oscillation along the dashed line in \figref{num}b is almost unity in the numerical calculation. It suggests it is possible to generate tunnel oscillation with high visibility, whereas the current oscillation visibility actually observed is pretty low (~3\%). We studied a dephasing problem as a possible origin for the low visibility. \figref{tun2}c shows comparison of the tunneling oscillation measured at 0.3 K and 7K. Here we used the left TCR and injected electrons from the upper wire. Weak, but similar features of current oscillations to those observed for the right TCR in \figref{tun2}b are identified as seen in \figref{tun2}d. But more importantly, there is no distinguishable change in the oscillation visibility with temperature. We note that 7K is the limit of measurable temperature in our setup. The only possible origin of the dephasing in our flying qubit is coupling to phonons. However, the phonon dephasing should depend on temperature, and therefore it seems negligible. This is probably because of the short dwell time of $\le 300$ ps during which the electron propagates through the TCR, and consistent with the theoretical calculation for charge qubits in a static double quantum dot previously reported by Thorwart et al.\cite{PhysRevB.72.235320}.

Then, what is the origin for the low visibility? This may only be assigned to poor fidelity of initialization of the electron wave function in the MQD entering the TCR. Electrons can be loaded directly to the excited states from the reservoirs. The potential roughness induced by dopant distribution in a depleted quantum wire can also scatter the electrons into the excited states. These processes can significantly disturb the inter-MQD tunneling but it is too complicated to consider them in the calculations.

In the numerical calculations, small dips of $P_{\rm out}$ are observed as indicated by yellow arrows (\figref{num}), although not visible in the experiment (See \figref{tun2}b). These dips originate from the tunnel coupling between the initially loaded ground state in the lower MQD and the higher excited states confined by the SAW potential along the traveling direction in the upper MQD. The tunnel coupling between these states is weak, because the corresponding wave functions are almost orthogonal to each other; thus, these minor dips only appear for more negative $V_{\rm gs}$, where the tunnel coupling is larger.

In summary, we observe coherent tunneling of SAW-driven single electrons between the two depleted but tunnel-coupled 1D wires. The coherent tunneling occurs when the MQDs are energetically aligned between the two wires with inter-channel tunneling strength and energy detuning as control parameters. The experimental data compares well to the numerical calculation. To the best of our knowledge, this is the first demonstration of coherent manipulation of a flying qubit encoded by SAW-driven single electron occupation in either of the two wires. This study is an important step toward realization of solid-state flying qubits.

We acknowledge fruitful discussion with Christopher B\"{a}uerle. 
M.Y. acknowledges support from KAKENHI (GrantNo.18H04284) and CREST-JST (No. JPMJCR1876). 
S.T. acknowledges support from CREST-JST (No. JPMJCR1675) and KAKENHI (GrantNo.18H04284).
A.L. and A.D.W. acknowledge gratefully support of DFG-TRR160,  BMBF - Q.Link.X  16KIS0867, and the DFH/UFA  CDFA-05-06.
The numerical calculations in this study were performed using the facilities at the Supercomputer Center, Institute for Solid State Physics, University of Tokyo.	
\bibliography{Papers.bib}
\end{document}

\newcommand{\beginsup}{%
        \setcounter{equation}{0}
        \renewcommand{\theequation}{S\arabic{equation}}%
        \setcounter{table}{0}
        \renewcommand{\thetable}{S\arabic{table}}%
        \setcounter{figure}{0}
        \renewcommand{\thefigure}{S\arabic{figure}}%
     }
\title{Supplemental Material: Coherent beam splitting of flying electrons driven by a surface acoustic wave}%

\author{R.Ito$^1$}
\author{S.Takada$^2$}
\author{A.Ludwig$^3$}
\author{A.D.Wieck$^3$}
\author{S.Tarucha$^1$}
\author{M.Yamamoto$^1$}
\affiliation{$^1$Center for Emergent Matter Science, RIKEN, 2-1 Hirosawa, Wako, Saitama 351-0198, Japan}
\affiliation{$^2$National Institute of Advanced Industrial Science and Technology, National Metrology Institute of Japan, 1-1-1 Umezono, Tsukuba,Ibaraki 305-8563,Japan}
\affiliation{$^3$Angewandte Festk\"{o}rperphysk, Ruhr-Universit\"{a}t Bochum, D-44780 Bochum Germany}

\maketitle

\beginsup

\section{Detail of numerical calculation}
\subsection{FEM}
We solved the Laplace equation by the finite element method (FEM) to calculate the potential profile defined by gate voltages. We performed the calculation for 5950 nm$\times$8940 nm in-plane area with 2 mm in height (perpendicular to the surface). Metal electrodes that define the tunnel coupled wires are placed around the center of the calculated area. In addition, top and bottom surface of the calculation space is defined as metals because ground metal plates are placed in these positions in the experiment. The electrical permittivity of the material under the metal electrodes is assumed to be that of GaAs, i.e. the relative permittivity of 13, and that of the rest is assumed to be that of vacuum. In the FEM calculation, the Dirichlet Boundary condition (fixed potential) is used for the surface of the gate electrodes and the Neumann boundary condition (no perpendicular electric field) is used for the calculation boundaries at the side surfaces.  The electrostatic influence of the two-dimensional electron gas (2DEG) which exists at 125nm below the surface is included in the calculation. Other sources to influence the potential profile such as dopant charges and surface charges are not included in the calculation. The dopant charges may shift the whole potential to the positive side and cause potential fluctuations. The surface charges should change the potential to the negative side. Note that charge accumulation of both dopants and surface charges is affected by the deposition of the gate electrodes. These may cause discrepancy of the gate voltages between the experiment and simulation.

The calculation is conducted by the following steps. First, we perform potential calculations, assuming that the 2DEG exists in the whole 2D area 125nm under the surface of the GaAs. Here, the 2DEG is treated as a zero potential boundary. Then, the charge accumulation on the 2DEG area is calculated. If the density of the positive accumulated charge exceeds the sheet electron density, the 2DEG element is changed to a GaAs element without 2DEG (The element is not assumed as zero potential boundary). The potential profile with the new configuration of the 2DEG element is then calculated. After repeating this sequence iteratively several times, the depletion region and potential profile of the entire sample is obtained. Note that the influence from the 2DEG to the potential profile is not so important because the qunantum wires are completely depleted by the gate voltage. It allows us to fix the 2DEG element area irrespectively to the gate voltages. Thus, each gate voltage linearly affects the electrostatic potential. 

\subsection{Simulation of electron transport}
We basically adopt the calculation method described in the paper by N.Watanabe et al. \cite{PhysRevE.62.2914}. First, the time evolution described by the Schr\"{o}dinger equation is approximated by separating the potential energy from the kinetic energy,
\begin{equation}
\psi(x,y,z,t+\Delta t)=e^{i\frac{\hbar\Delta t}{4m}\frac{\partial^2}{\partial x^2}}e^{i\frac{\hbar\Delta t}{4m}\frac{\partial^2}{\partial y^2}}e^{-i\frac{\Delta t V(x,t+\frac{\Delta t}{2})}{\hbar}}e^{i\frac{\hbar\Delta t}{4m}\frac{\partial^2}{\partial y^2}}e^{i\frac{\hbar\Delta t}{4m}\frac{\partial^2}{\partial x^2}}\psi(x,y,z,t). 
\end{equation}
$\Delta t$ is the time duration of each step in the finite difference method. The calculation of each exponential is conducted step by step. For example to calculate $\psi^{\prime}(x,t)=e^{i\frac{\hbar\Delta t}{4m}\frac{\partial^2}{\partial x^2}}\psi(x,t)$, the equation is first decomposed into $e^{-i\frac{\hbar\Delta t}{8m}\frac{\partial^2}{\partial x^2}}\psi^{\prime}(x,t)=e^{i\frac{\hbar\Delta t}{8m}\frac{\partial^2}{\partial x^2}}\psi(x,t)$ and then the first order Taylor expansion is performed. 
\begin{equation}
(1-i\frac{\hbar\Delta t}{8m}\frac{\partial^2}{\partial x^2})\psi^{\prime}(x,t)=(1+i\frac{\hbar\Delta t}{8m}\frac{\partial^2}{\partial x^2})\psi(x,t)+\mathcal{O}(\Delta t^2).
\label{aa}
\end{equation}
Note that this approximation keeps Unitarity.  Now the wave function is discretized for $x$. The differential term is given as $\frac{\partial^2}{\partial x^2}\psi(x_i,t)=\frac{\psi(x_{i+1},t)-2\psi(x_i,t)+\psi(x_{i-1},t)}{\Delta^2 x}$ with $\Delta x =x_i-x_{i-1}$. The problem is thus reduced to solving simple linear algebra. In our calculation, we chose $\Delta x$ of 10nm with corresponding times $\Delta t$ of $\sim$0.03ps. $V(x,t)$ consists of the electrical potential calculated by FEM and a SAW potential defined as a 20mV amplitude sinusoidal electrical potential wave. We assumed that these potentials do not interact with each other.

\section{Additional experimental result}
\subsection{Total current dependence}
\begin{figure}
	\begin{center}
		\includegraphics[width=10cm]{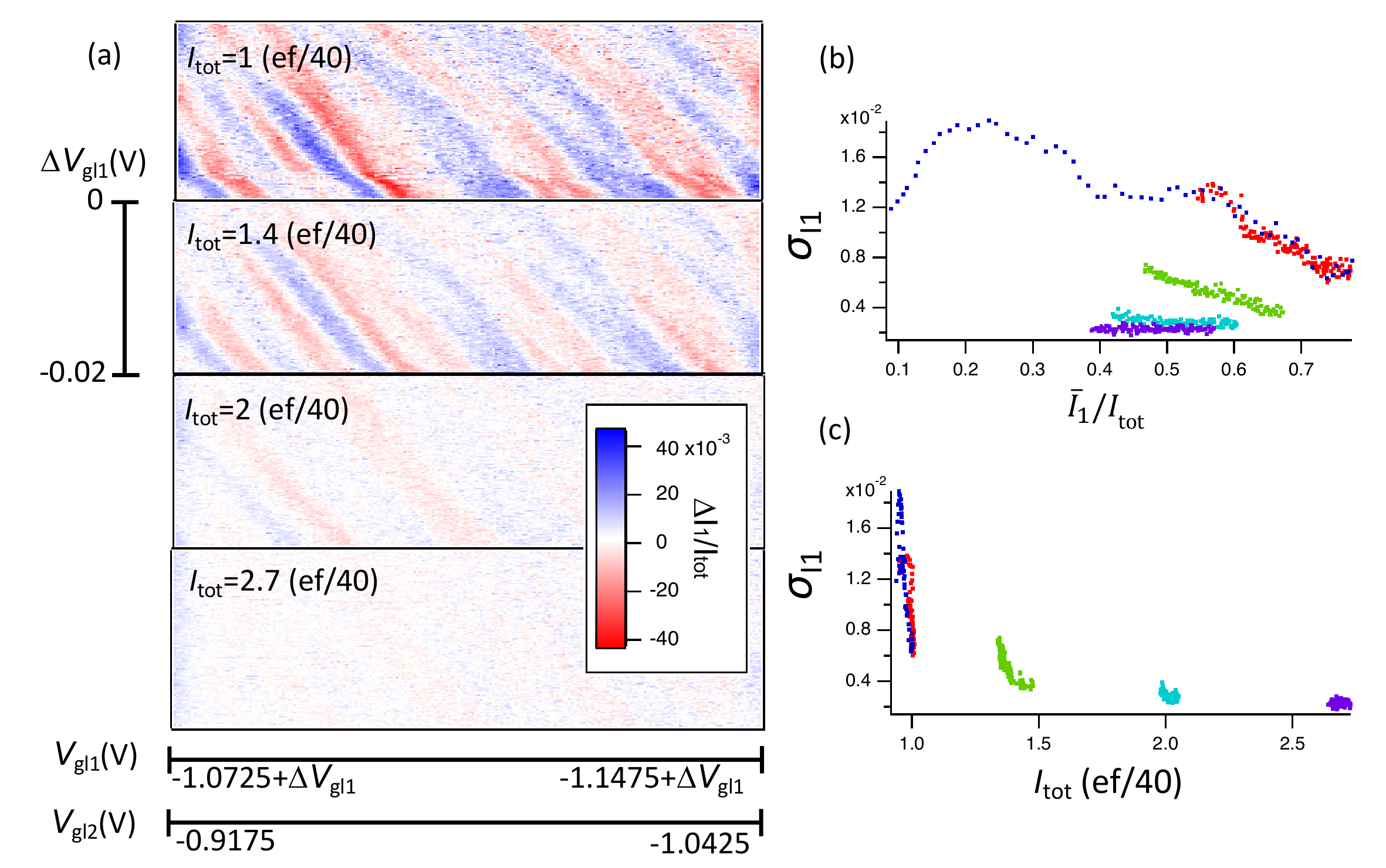}
		\caption{(a) Oscillation patterns for different numbers of injected electrons in each SAW potential minimum. The oscillation component divided by the total current is displayed. (b) and (c) Standard deviation of the oscillation component as a function of the normalized current $\overline{I_1}/I_{\rm tot}$ and total current $I_{\rm tot}$, respectively. The data sets obtained for different $I_{\rm tot}$ are shown with different colors. \label{th}} 
	\end{center}
\end{figure}

To confirm the scenario of coherent tunneling of a single electron, we investigated the oscillation for different numbers of electrons in each MQD as shown in \figref{th}a. In this measurement, the left hand side TCR in \figref{dev2}a is used and electrons are injected from the upper wire. The number of electrons in each MQD is varied from 1 to 2.7, using the entrance gate voltage $V_{\rm e1}$. While the direct capacitive coupling between the entrance gate and the TCR is sufficiently small, the current oscillation in total current, $\Delta I_1/I_{\rm tot}$, is significantly suppressed by increasing the number of electrons. Clear oscillation is only observed when a single electron is trapped in each MQD. 

The gate voltage in \figref{th}a is tuned such that the non-oscillationg components of the current ($\overline{I_i}=I_{\rm i} - \Delta I_{\rm i}$, i=1,2) are constant along the horizontal direction. \figref{th}b and c show the standard deviation of the oscillation component ($\sigma_{\rm I1}$), evaluated by calculating the root mean square of $\Delta I_{\rm 1}/I_{\rm tot}$ along the horizontal direction of \figref{th}a. The standard deviation is shown as a function of normalized current, $\overline{I_{1}}/I_{\rm tot}$ and $I_{\rm tot}$ in \figref{th}b and c, respectively. The colors of points in these figures indicate $I_{\rm tot}$. Red, green, cyan and purple points belong to the groups with $I_{\rm tot}$ of 1,1.4,2 and 2.7 ef/40 respectively as shown in \figref{th}c and they are derived from the data set in \figref{th}a. Blue points are derived from another data set. In \figref{th}c,  $\sigma_{I1}$ clearly decreases as $I_{\rm tot}$ increases. In \figref{th}b, $\sigma_{I1}$ shows a clear dependence on $\overline{I_1}/I_{\rm tot}$. For the smaller $\overline{I_1}/I_{\rm tot}$, the lower energy states in the upper MQD plays a main role for the tunneling oscillation. The peak of $\sigma_{I1}$ at small $\overline{I_1}/I_{\rm tot}$ appears due to the energy distribution of injected electrons. 

\subsection{Electron drop}
\begin{figure}
\begin{center}
	\includegraphics[width=12cm]{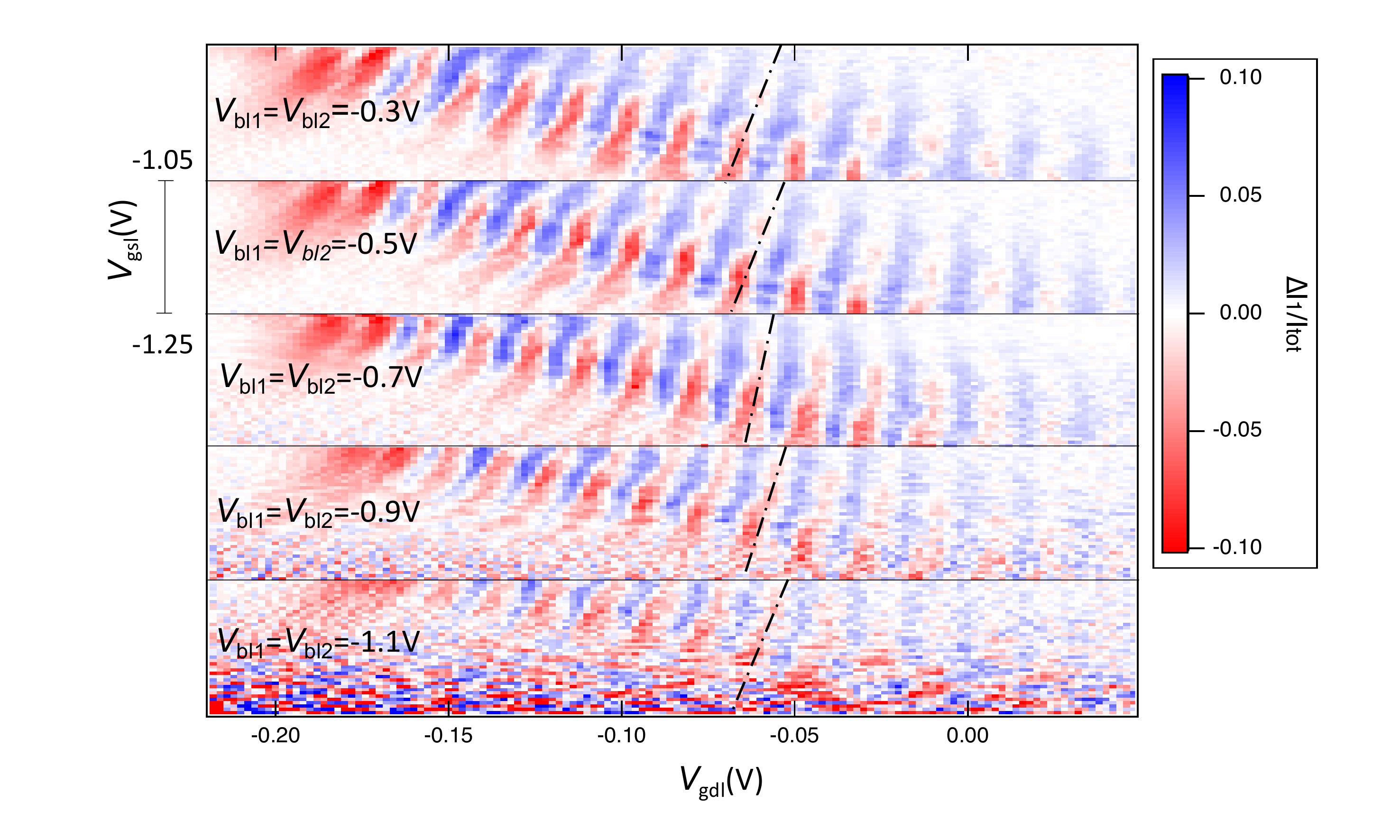}
	\caption{Current oscillation signal observed for the left TCR. $V_{\rm bl1}$ and $V_{\rm bl2}$ are modified to control the potential at the exit of the coupled region.\label{backgate}}
\end{center}
\end{figure}

\figref{backgate} shows the current oscillation data with $V_{\rm bl1}$ and $V_{\rm bl2}$ as a parameter. The lefthand-side TCR is used for the measurement and single electrons are injected from the upper channel. The color plot shows the normalized oscillation component. When the patterns of different $V_{\rm bl1}$ and $V_{\rm bl2}$ are compared, we find that the visibility of these patterns is different. At higher $V_{\rm bl1}$ and $V_{\rm bl2}$, we see a clear signal of the coherent oscillation (along the dot-dashed line), but it becomes invisible for lower $V_{\rm bl1}$ and $V_{\rm bl2}$. This is also reproduced by the numerical calculation depicted in \figref{num} (where we discuss the drop of an electron from a MQD). This supplemental result is  consistent with the calculation result, where a clearer oscillation pattern is observed when electrons drop from the SAW confinement. 

In the experiment, the total current $I_{\rm tot}$ is not adjusted and decreases from ef/40 to 0.2 ef/40 at $V_{\rm bl1}= V_{\rm bl2}=-0.3$V and $V_{\rm bl1}= V_{\rm bl2}=-1.1$V,  respectively. Note that $I_{\rm tot}$ is also dependent on $V_{\rm gsl}$. $I_{\rm tot}$ is 0.4 ef/40 at $V_{\rm gsl}= -1.05$ V and zero at $V_{\rm gsl}= -1.25$ V for $V_{\rm bl1}= V_{\rm bl2}=-1.1$V, respectively. The reduction of the total current lowers the signal-to-noise ratio.  
\bibliographystyle{unsrt.bst}
\bibliography{Papers}